# Polymorph-Tunable Spin Texture and Excitonic Structures in Monolayer $WSi_2P_4$

Xianbo Chenwei[1], Yu Zhou[1], Ke Wu[1], Gaofeng Xu[1], Ruixue Li[1], Yuan Li[1], Yabei Wu[2*], Shaowen Xu[3*], and Fanhao Jia[1*]

[1]Department of Physics, Hangzhou Dianzi University, Hangzhou 310018, China

[2]Department of Materials Science and Engineering, Southern University of Science and Technology, Shenzhen, Guangdong 518055, China

[3]School of Physics and Optoelectronic Engineering, Hangzhou Institute for Advanced Study, University of Chinese Academy of Sciences, Hangzhou 310024, China

[*]Contact author: fanhaojia@shu.edu.cn
[*]Contact author: xushaowen@ucas.ac.cn
[*]Contact author: wuyb3@sustech.edu.cn

## ABSTRACT

Two-dimensional semiconductors that simultaneously possess a direct band gap and strong spin–orbit coupling (SOC) are highly attractive for quantum optoelectronics. Using first-principles GW plus Bethe–Salpeter equation (GW-BSE) calculations, we show that monolayer $WSi_2P_4$—an experimentally accessible member of the $MSi_2X_4$ family—hosts a direct K-valley gap together with strong SOC. Its three competing polymorphs ($\alpha$, $\beta$, $\gamma$) are all direct-gap semiconductors and are kinetically locked behind ~1.3 eV migration barriers. The $\sigma_h$ mirror plane of the $D_{3h}$ of $\alpha$ and $\gamma$ phases enforces a persistent spin texture across the Brillouin zone, weakly modulated near Γ by interband SOC mixing, whereas the polar $\beta$ phase ($C_{3v}$) exhibits Rashba spin splitting. SOC splits the doubly-degenerate lowest bright exciton into dark states and splits the original absorption peak into two peaks (A and B), whose relative brightness is governed by the K-valley conduction-band splitting $\Delta_{CB}$. $\alpha$-$WSi_2P_4$ displays the brightest Peak A because a tiny band crossing of opposite-spin branches can open a spin-allowed radiative channel. These results establish the $MSi_2P_4$ family as a phase-tunable platform for the cooperative engineering of spin texture, band splitting, and excitonic brightness within a single material system.

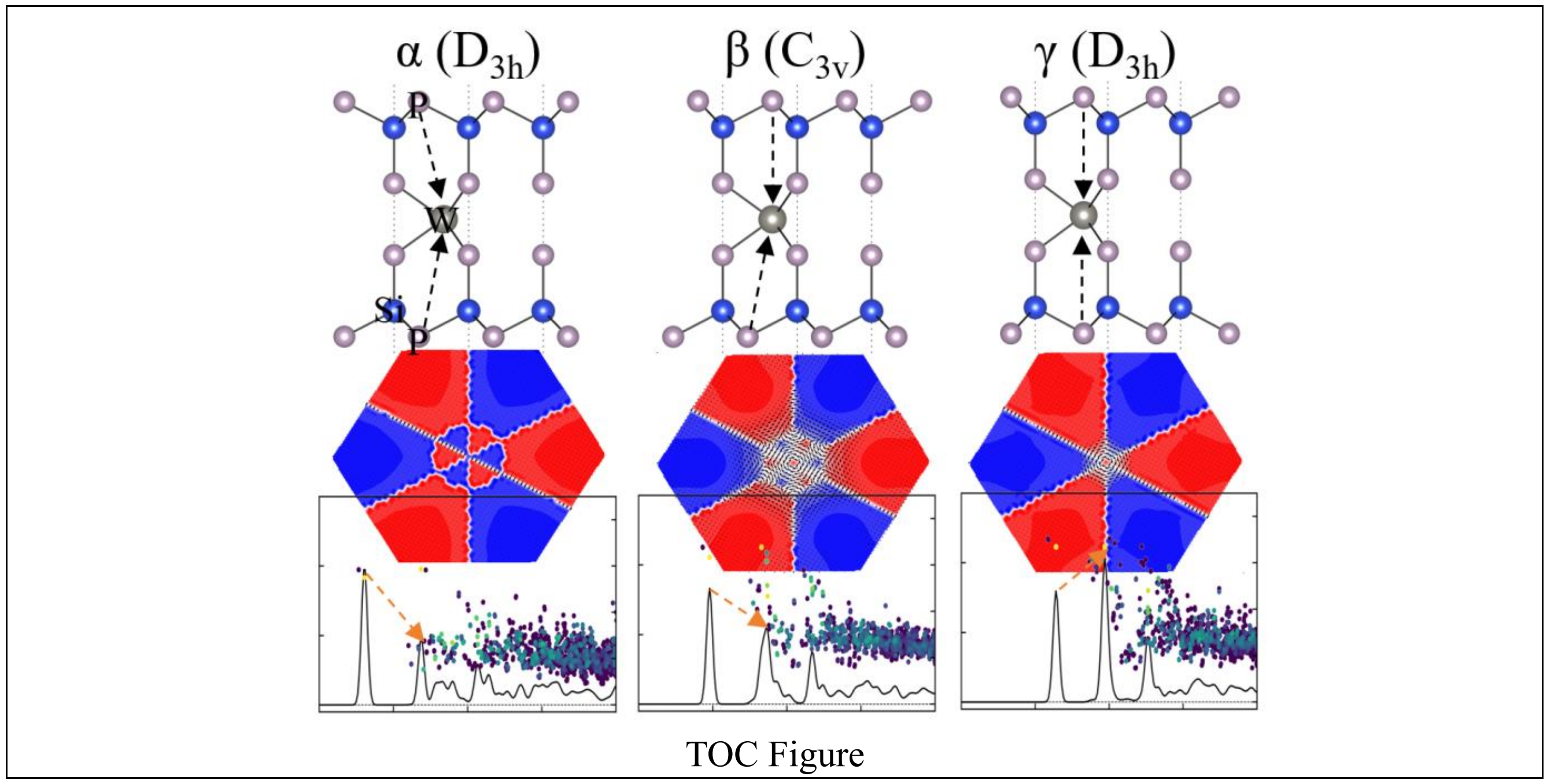

TOC Figure

Two-dimensional (2D) van der Waals (vdW) semiconductors are central to post-Moore electronics[1] and quantum optoelectronics.[2] A major advance in this field occurred in 2020,[3] when $MoSi_2N_4$ was synthesized by passivating the surface of a 2D non-layered molybdenum nitride with elemental silicon, thereby opening a new vdW family with the general formula $MSi_2X_4$ (M = Mo, W; X = N, P, As, Sb).[4-5] Its septuple-layer architecture (X–Si–X–M–X–Si–X) has no naturally occurring layered bulk counterpart and has been obtained by bottom-up chemical vapor deposition. This unique encapsulation endows the $MSi_2X_4$ family with exceptional structural stability, protected band edge states,[6] and rich structural and compositional tunability—including phase polymorphism and site-specific anion substitution—making it an attractive complementary platform for functional 2D materials.[7] Very recently, wafer-scale epitaxial growth of *p*-type monolayer $MoSi_2N_4$ single crystals was achieved on Cu(111) substrates, yielding intrinsic carrier mobilities as high as 154 $cm^2V^{-1}s^{-1}$ and excellent device stability, which demonstrates the technological viability of this material family.[8]

A central challenge for the $MSi_2X_4$ family, however, lies in its electronic structure. The experimentally realized $MoSi_2N_4$ and $WSi_2N_4$ monolayers are both indirect-bandgap semiconductors,[6, 9] a fact that severely limits their optoelectronic performance.[10] A promising strategy to overcome this limitation is to enhance the p-d hybridization by replacing N with heavier pnictogens (P, As, Sb).[11] Our previous theoretical study has demonstrated that such substitution not only drives a ground-state transition from the $\alpha$ to the $\gamma$ phase, but, more importantly, converts the band gap from indirect to direct at the K point—a feature that holds across all three polymorphs ($\alpha$, $\beta$, $\gamma$).[12] This opens a viable pathway to direct-gap $MSi_2X_4$ semiconductors that simultaneously combine strong spin–orbit coupling (SOC) arising from the heavy transition-metal atom, thereby creating new opportunities for high-efficiency optoelectronic[13-14] and spintronic applications.[15]

Among the X-substituted members, $MSi_2P_4$ stands out as a particularly intriguing system. While theoretical studies predict that the nitride counterparts ($MSi_2N_4$) also possess three energetically close polymorphs ($\alpha$, $\beta$, and $\gamma$) separated by large kinetic barriers of ~1.8 eV,[12] experiments have so far realized only the $\alpha$ phase, which is the ground state. In contrast, for $WSi_2P_4$, the $\gamma$ phase is the energetic ground state—a key distinction from the nitrides. Moreover, the energy differences among the $\alpha$, $\beta$, and $\gamma$ phases of $MSi_2P_4$ are remarkably smaller, with a maximum of only ~10 meV/atom, compared with the ~25 meV/atom found in $MSi_2N_4$. This suggests that multiple polymorphs of $MSi_2P_4$ may be experimentally accessible. All three phases of $MSi_2P_4$ are direct-bandgap semiconductors with quasiparticle gaps of 0.69, 0.98, and 1.18 eV, respectively, and they are separated by lower yet still substantial kinetic barriers of approximately 1.3 eV, ensuring that each phase can persist as a pure polymorph without facile interconversion. In addition, all three phases satisfy dynamical stability criteria, and their distinct point-group symmetries—$D_{3h}$ for $\alpha$ and $\gamma$, and $C_{3v}$ for the polar $\beta$ phase—give rise to markedly different spin textures and band-splitting characteristics. The $\sigma_h$ mirror plane of the $D_{3h}$ phases enforces a persistent spin texture (PST) over the full Brillouin zone (full-BZ)[16]—a rare situation that protects spin information against spin-precession decay.[17-19] Moreover, the absence of inversion symmetry permits higher-order interband SOC mixing,[20] which relaxes the strict collinearity of the spin texture and yields in-plane spin components reminiscent of the Rashba texture of the polar $\beta$ phase. These symmetry-governed spin textures, combined with the strong SOC of W, are expected to profoundly influence excitonic properties through mechanisms such as spin-forbidden transitions, momentum mismatch, and density-of-states enhancement.[21-22]

With the growing interest in the $MSi_2X_4$ family, a comprehensive understanding of how spin texture, band splitting, and excitonic brightness vary across polymorphs is highly desired.[23] In particular, the interplay between the symmetry-enforced persistent spin texture and the SOC-induced splitting of band edges at the K valley—the region that governs the optical response—has not been systematically explored. Here, using $WSi_2P_4$ as a prototypical direct-gap member of the $MSi_2X_4$ family, we perform first-principles GW-BSE calculations to investigate the excitonic structure and spin texture of the $\alpha$, $\beta$, and $\gamma$ monolayers. The density functional theory (DFT) calculations were performed using the Vienna *ab initio* simulation package (VASP),[24] while GW[25] and BSE[26] calculations were performed with BerkeleyGW[27] using Kohn–Sham wave functions from Quantum ESPRESSO[28] with fully relativistic optimized norm-conserving (NC) Vanderbilt pseudopotentials[29] obtained from the PseudoDojo repository.[30] We demonstrate that the three phases exhibit distinct excitonic brightness patterns dictated by the conduction-band splitting at K, with $\alpha$-$WSi_2P_4$ showing the brightest low-energy exciton owing to a spin-allowed radiative channel related to the tiny band crossing. Our results establish $WSi_2P_4$ as a platform in which spin texture, band splitting, and excitonic brightness can be jointly engineered, and suggest that phase selection offers a practical handle for tuning spin-resolved emission in the infrared.

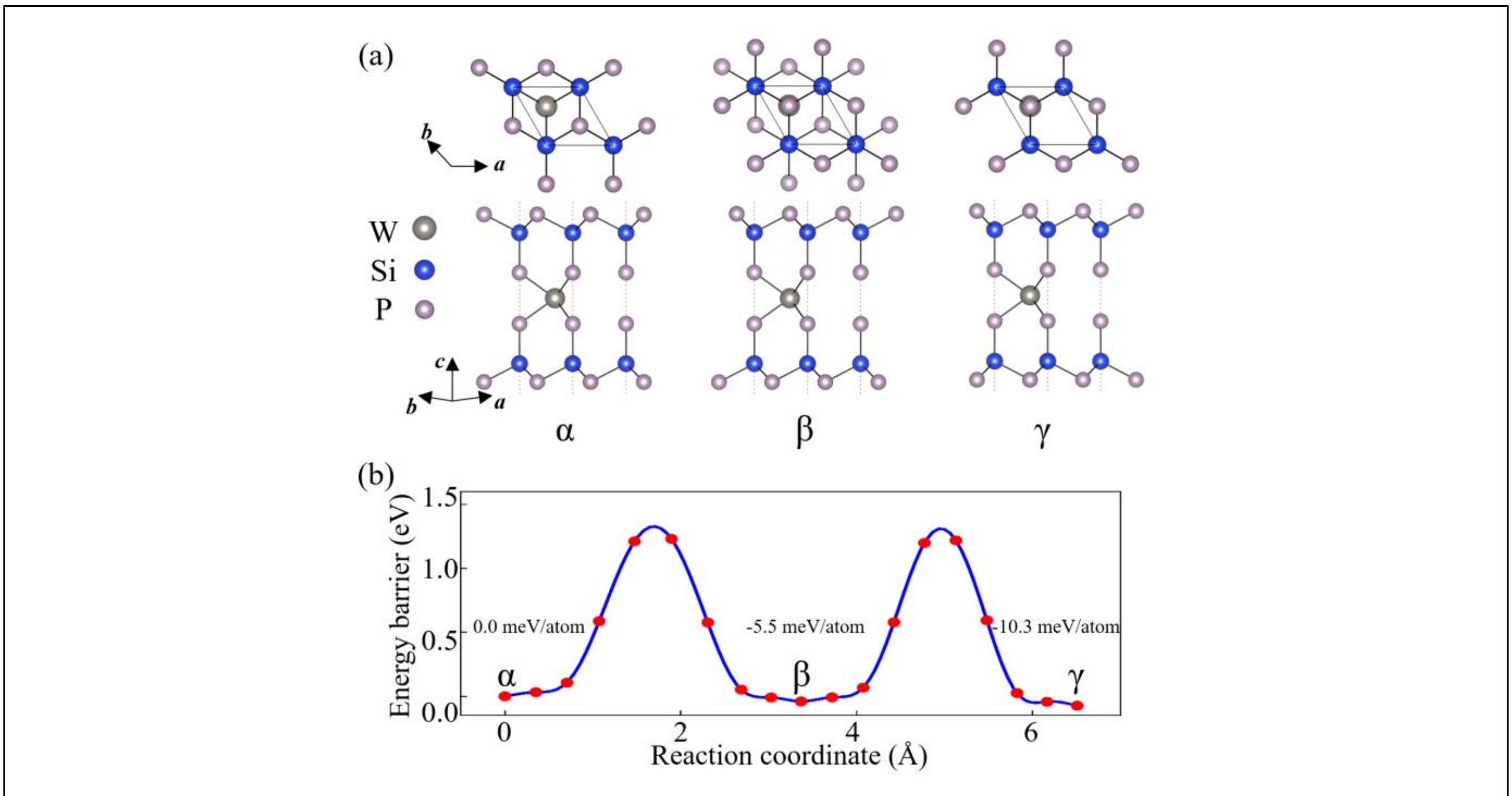

Figure 1. (a) Top and side views of the crystal structures of $WSi_2P_4$ monolayers: $\alpha$, $\beta$, and $\gamma$ phases. (b) NEB energy barriers for the migration of the P atom of the top atomic layer along the path from $\alpha$ to $\beta$, and to γ.

Figure 1a illustrates the atomic geometries of the α, β, and γ phases of monolayer $WSi_2P_4$. The $\alpha$ and $\gamma$ phases share the same point group ($D_{3h}$) and both possess an in-plane mirror plane $\sigma_h$. The sole distinction between them lies in the Wyckoff occupancy of the outer P layers bonded to Si: in the $\alpha$ phase, the outer P atoms occupy the 2i site and are staggered by 60° relative to W, whereas in the $\gamma$ phase they occupy the 2h site and are aligned with W. The remaining atoms are identical in both phases: W at 1d (1/3, 2/3, 1/2), and Si and inner P at 2g (0, 0, ±z). The $\beta$ phase differs fundamentally from $\alpha$ and $\gamma$: it is polar, with the top and bottom outer P layers residing on different columns (bottom at Wyckoff 1c, top at 1b), which removes the $\sigma_h$ mirror plane and lowers the point group from $D_{3h}$ to $C_{3v}$. The optimized lattice parameters are summarized in Table 1. Energetically, the three phases are nearly degenerate: $\beta$ lies 5.5 meV/atom below $\alpha$, and $\gamma$ is the ground state, lying 10.3 meV/atom below $\alpha$. Figure 1b presents the nudged elastic band (NEB) energy profile for the $\alpha \rightarrow \beta \rightarrow \gamma$ transformation. Despite the small energy differences, both transitions must overcome barriers of ~1.3 eV, indicating that each phase can persist as a pure polymorph without facile interconversion. Notably, this barrier is considerably lower than the ~1.8 eV barriers calculated for $MSi_2N_4$,[12] suggesting that $WSi_2P_4$ is more prone to polymorph formation than its nitride counterparts. Depending on the specific synthesis pathway, monolayer $WSi_2P_4$ may therefore exist as a mixture of polymorphs. Our previous study has confirmed that all three phases satisfy dynamical and thermodynamic stability criteria.[12]

**Table 1.** The optimized lattice constant *a*, monolayer thickness *d*, relative energies ΔE (meV/atom) and band gaps ($E_g$, in eV) of *α*, *β*, and γ phases of $WSi_2P_4$ monolayers at the PBE, PBE+SOC, GW, and GW+SOC levels.

| | $a$ (Å) | $d$ (Å) | ΔE | $E_g^{PBE}$ | $E_g^{PBE+SOC}$ | $E_g^{GW}$ | $E_g^{GW+SOC}$ |
|---|---|---|---|---|---|---|---|
| $\alpha$ | 3.437 | 9.335 | 0.0 | 0.55 | 0.32 | 1.01 | 0.69 |
| $\beta$ | 3.429 | 9.377 | -5.5 | 0.77 | 0.54 | 1.25 | 0.98 |
| $\gamma$ | 3.420 | 9.420 | -10.3 | 0.94 | 0.69 | 1.44 | 1.18 |

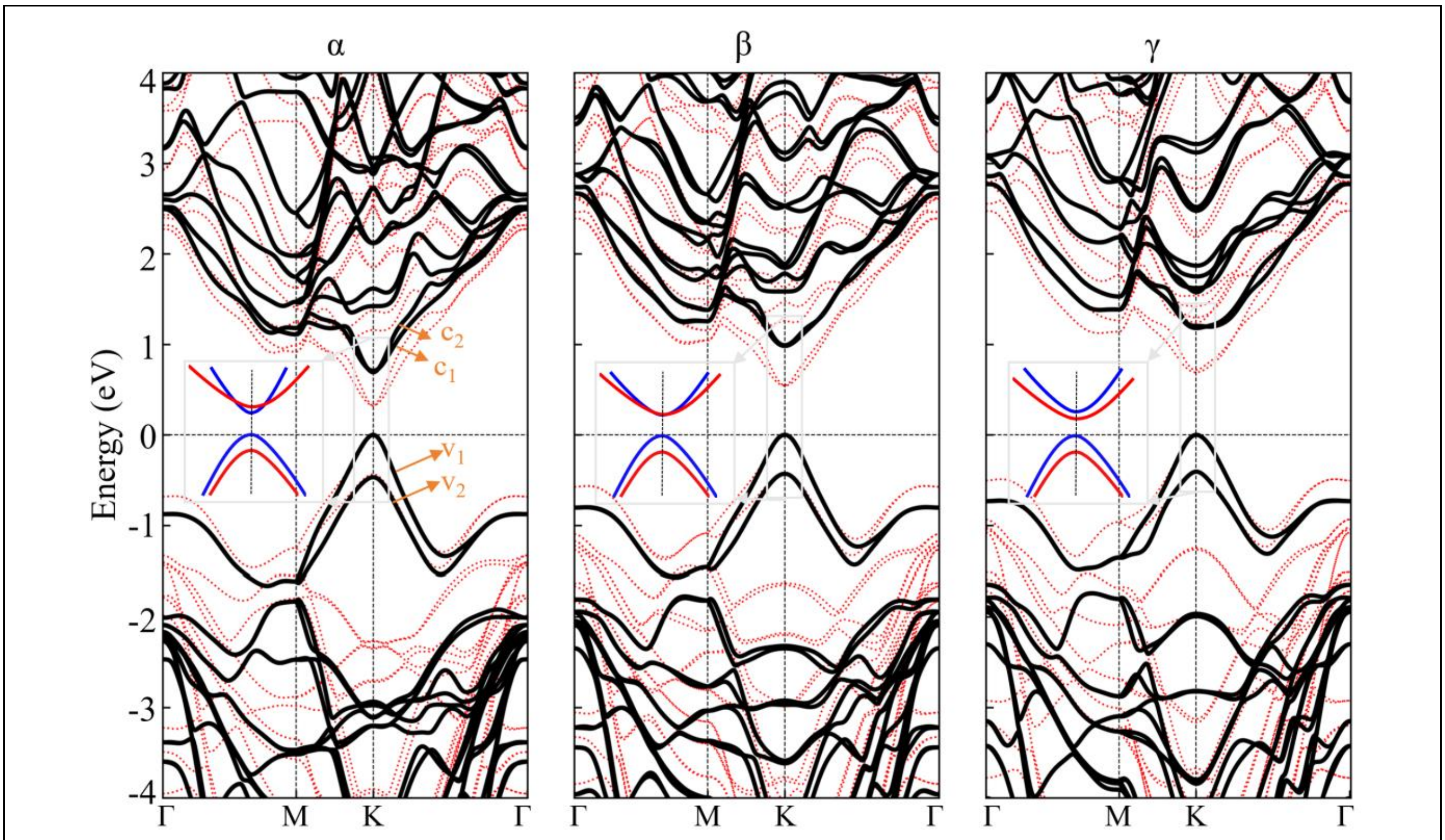

Figure 2. Band structures of three $WSi_2P_4$ monolayers. The red dashed and black solid lines were calculated using PBE+SOC and GW+SOC, respectively. Insets are the schematic diagram of the spin splitting of band edges around K. Red/blue lines denote spin-up/spin-down $S_z$ components.

Figure 2 shows the band structures of the three $WSi_2P_4$ monolayers calculated at the PBE+SOC and GW+SOC levels. In all cases, both the CBM and VBM are located at the K point, confirming their direct-gap character. The band structures without SOC are provided in Figure S1. The GW+SOC quasiparticle gaps are 0.69, 0.98, and 1.18 eV for the $\alpha$, $\beta$, and $\gamma$ phases, respectively, placing the optical response of $WSi_2P_4$ in the infrared spectral range for optoelectronic applications. Projected band analysis (Figures S2 and S3) reveals that the VBM at K is dominated by W $d_{xy}$ and $d_{x^2-y^2}$ orbitals with minor P $p_x$/$p_y$ contributions, whereas the CBM at K is primarily W $d_{z^2}$. At Γ, the VBM is governed by W $d_{z^2}$–P $p_z$ hybridization, while the CBM arises from P $p_{x/y}$ hybridized with W $d_{yz/xz}$. Notably, although $\alpha$ and $\gamma$ share the same point group and differ only in the stacking registry of the outer P atoms (staggered vs. aligned with respect to W), their band gaps differ by nearly 70%. This disparity results from two cooperating orbital-hybridization effects. For the CBM, the relevant orbital pair is W $d_{z^2} \leftrightarrow$ P $p_z$, corresponding to a vertical σ bond (antibonding state). In the $\gamma$ phase, the aligned geometry maximizes overlap, significantly raising the CBM. For the VBM, the orbital pair is W $d_{xy/x^2-y^2} \rightarrow$ P $p_{x/y}$, corresponding to in-plane bonding states. The aligned geometry in $\gamma$ enhances the in-plane ligand field, substantially lowering the VBM. Consequently, the gap increases from 0.69 eV ($\alpha$) to 1.18 eV ($\gamma$) as a result of simultaneous CBM elevation and VBM depression. The strong SOC arising from W is evident in the band structures. PBE predicts SOC-induced bandgap modifications of 0.23–0.25 eV, while GW yields 0.26–0.32 eV. Given that all three phases lack inversion symmetry—and $\beta$ is additionally polar—the SOC gives rise to diverse spin-splitting phenomena and nontrivial spin textures (e.g., Rashba effects[31]) that are expected to profoundly influence excitonic properties through mechanisms such as spin-forbidden transitions, momentum mismatch, and density-of-states enhancement.

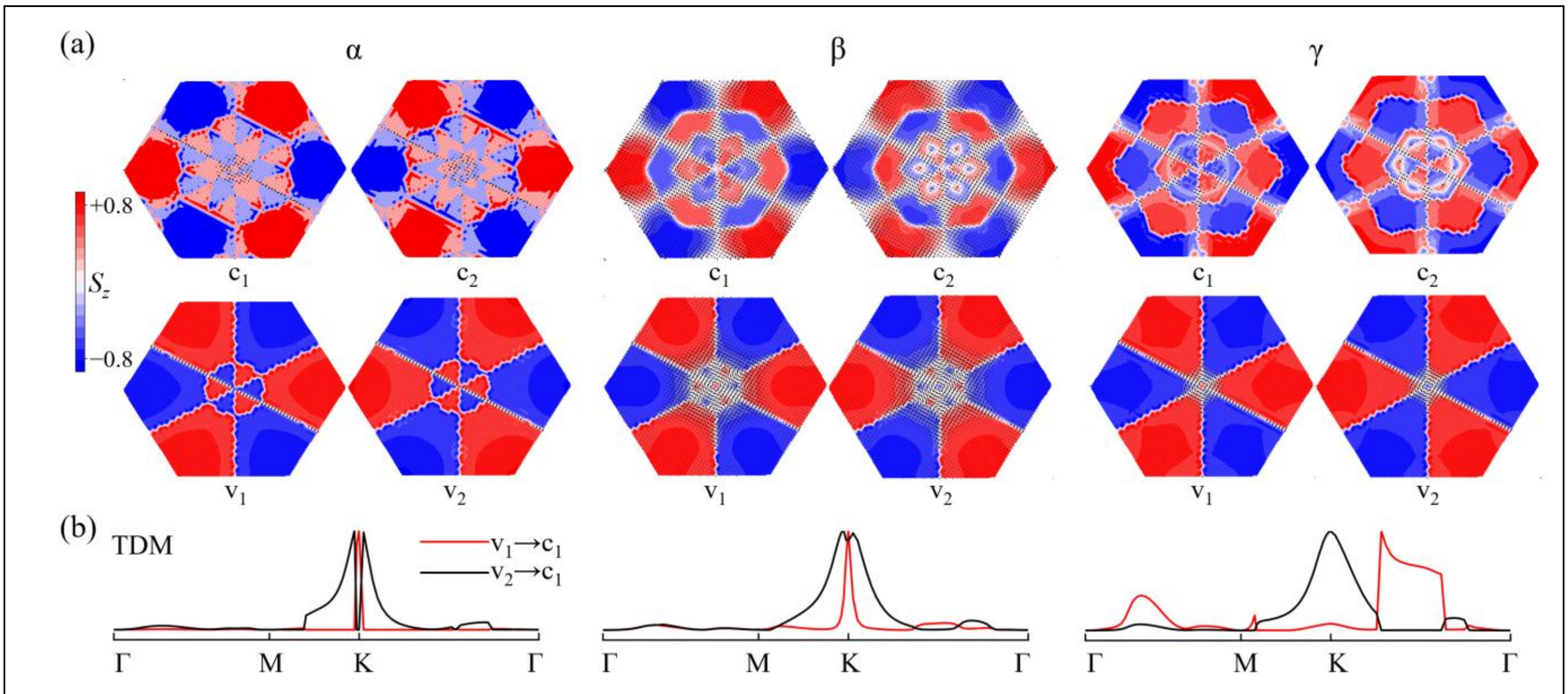

Figure 3. (a) Spin textures of the top two valence bands and bottom two conduction bands of three $WSi_2P_4$ monolayers, where $v_1$, $v_2$, $c_1$, and $c_2$ were defined in Figure 2. The black arrows provide the in-plane spin component, while the colors indicate the out-of-plane component of spins: red is spin-up and blue is spin-down. (b) Normalized transition dipole moment along the high-symmetry band path.

We first examine the spin texture. Conventional 2D Rashba spin textures typically emerge near time-reversal-invariant momenta (TRIM), such as Γ. However, the band edges of all three $WSi_2P_4$ phases reside at K, which is a non-TRIM point. At K, a zeroth-order ($k$-independent) $\sigma_z$ splitting—the Ising (spin-valley) coupling[32]—is permitted, whereas it is forbidden at Γ by time-reversal symmetry. For the $\alpha$ and $\gamma$ phases (point group $D_{3h}$), the absence of inversion symmetry allows a nonzero spin polarization $P(k) = \langle\psi|\hat{S}|\psi\rangle$, while the in-plane mirror plane $\sigma_h$ constrains its direction. Consider a nondegenerate band at a general $k$ point; its wavefunction must be an eigenstate of $\sigma_h$. Under $\sigma_h$, $S_x \rightarrow -S_x$ and $S_y \rightarrow -S_y$, which enforces $\langle S_x \rangle = \langle S_y \rangle = 0$. Therefore, only the $z$-component of spin polarization is allowed: $P(k) = P_z(k)\hat{z}$. Time-reversal symmetry further requires $P_z(-k) = -P_z(k)$, which dictates the sign reversal of the spin texture in momentum space (Figure 3) without altering its orientation. This is the hallmark of a persistent spin texture (PST): the spin polarization is everywhere perpendicular to the plane, varying only in magnitude across the Brillouin zone. Sheoran et al. showed that $WSi_2N_4$ satisfies the conditions for full-BZ PST. [16]

Near K, the SOC splitting is large (dominated by the zeroth-order Ising term Δ), and the bands are nondegenerate. Consequently, the wavefunctions remain approximately $\sigma_h$ eigenstates, preserving an almost perfect out-of-plane PST. In the nondegenerate limit, interband coupling is suppressed by the large band separation, so the perturbed states retain their out-of-plane spin character. However, when bands become degenerate or nearly degenerate—as occurs near Γ—this picture can break down. In the $\gamma$ phase, we observe a pronounced *Rashba-like* in-plane spin texture near Γ. This arises because the SOC Hamiltonian couples between even-parity and odd-parity bands. Such coupling can be remote and does not violate the $\sigma_h$ symmetry of the system. Since $S_{x/y}$ are odd under $\sigma_h$ while the even- and odd-parity wave functions have opposite parity, the integrand $\langle \text{even}|S_{x/y}|\text{odd}\rangle$ is even and does not vanish. This mechanism is known as $k$-mediated interband SOC mixing.[33-34] The intrinsic $\sigma_h$ symmetry forbids a conventional (first-order) Rashba term in $\alpha$ and $\gamma$; nevertheless, strong k·p-mediated remote interband SOC mixing (second-order perturbation in $H_{SO}$) generates in-plane spin components near Γ. The mixing strength is governed by an effective coupling coefficient[35-36]: $\eta_{\text{eff}} \propto \frac{\lambda \cdot v_F \cdot k}{\Delta E}$, where λ is the SOC strength of the W 5d orbitals (~0.1–0.4 eV for W), ΔE is the energy difference between the even-parity VBM and the odd-parity remote band, and $v_F$ is the k·p velocity matrix element $\langle \text{even} \mid \partial/\partial_x \mid \text{odd}\rangle$, which varies slightly with geometry but remains of similar magnitude. For the $\gamma$ phase (P−W−P aligned), the remote interband mixing near Γ involves the even-parity top valence band (W $d_{z^2}$ +P $p_z$ hybrid) and the odd-parity band approximately 0.41 eV below (W $d_{xz/yz}$ + P $p_{x/y}$ hybrid). With ΔE = 0.41 eV, the mixing factor λ/ΔE ≈ 0.24 (taking λ = 0.1 eV), indicating that the mixing is strong enough to produce clearly resolved in-plane components $\langle S_{x/y}\rangle$. In contrast, for the $\alpha$ phase (P-W-P staggered), the top valence band has ΔE = 0.63 eV, giving λ/ΔE ≈ 0.16 is substantially smaller; consequently, the in-plane components are largely suppressed. For the bottom conduction band near Γ, the situation reverses. The $\alpha$ phase has a much smaller ΔE = 0.14 eV between the odd-parity ($d_{xz/yz}$) and even-parity ($d_{xy/x^2-y^2}$) orbitals than the γ phase (0.29 eV). Consequently, the α phase exhibits pronounced in-plane spin components $\langle S_{x/y}\rangle$ near Γ in the bottom conduction band, substantially exceeding those in the γ phase.

We now turn to SOC-induced band splitting. Despite the pronounced in-plane spin textures near Γ discussed above, the SOC-induced splitting of both the VB and CB remains weak in all three phases. This reflects the predominant P p-orbital character of the band edges near Γ: the weak atomic SOC and small orbital angular momentum of the P p states limit the splitting in all three phases, even though $\beta$ exhibits a conventional Rashba-type texture. In contrast, the Zeeman-like splitting at the K valley is strong. Since the band-edge electronic states governing the optical properties of $WSi_2P_4$ reside primarily near K, we focus on the splitting characteristics there. Low-energy excitons are determined by the two lowest CBM bands ($c_1$, $c_2$) and two topmost VBM bands ($v_1$, $v_2$) near K (defined in Figure 2). The $c_1$ and $c_2$ bands are dominated by $d_{z^2}$ (orbital angular momentum $m_l =$ 0), while $v_1$ and $v_2$ are dominated by $d_{xy/x^2-y^2}$ ($m_l = \pm 2$).

The insets of Figure 2 illustrate the band splitting near the CBM and VBM. From the spin ordering of the band edges, spin matching dictates that the $v_1 \rightarrow c_1$ transition is forbidden at K in β and γ, while $v_2 \rightarrow c_1$ is allowed, as confirmed by the transition dipole moment calculations in Figure 3b. The valence-band splittings ΔVB at K exceed 400 meV in all three phases (434.5, 417.3, and 415.7 meV; Table S1) and are Zeeman-like. This arises because the $\lambda L_z S_z$ term in the SOC Hamiltonian $H_{SO}$ acts on $m_l = \pm 2$ states, producing a large splitting of $2\lambda \cdot |m_l| \approx 4\lambda$. Taking $\lambda \approx 0.1$ eV yields values in good agreement with DFT. For the CBM, since $d_{z^2}$ has $m_l = 0$, the first-order SOC splitting $\Delta_{CB}$ should vanish. However, $\Delta_{CB}$ is nonzero in our DFT calculations (though much smaller than $\Delta_{VB}$). This nonzero ΔCB arises from second-order perturbative coupling mediated by remote bands (e.g., $d_{xy}$) or from hybridization. For the $\alpha$ phase, negative $\Delta_{CB} = -6.0$ meV implies a band crossing of the spin-split conduction bands. This crossing only happens within a tiny region (~0.01 × 2π/a) near K, but it may play a crucial role in the excitonic properties. This crossing originates from the difference in effective masses between spin-up and spin-down channels. Since both spin channels are $\sigma_h$ eigenstates and are fully decoupled, the energy eigenvalues are:

$$E_{\uparrow}(k) = \frac{\hbar^2 k^2}{2m_{\uparrow}^*} + \Delta,\ E_{\downarrow}(k) = \frac{\hbar^2 k^2}{2m_{\downarrow}^*} - \Delta.$$

When $m_{\uparrow}^* \neq m_{\downarrow}^*$ and Δ is a small constant of a few meV, the two parabolas have different slopes and must cross at some finite momentum $k_0$:

$$\frac{\hbar^2 k_0^2}{2m_{\uparrow}^*} + \Delta = \frac{\hbar^2 k_0^2}{2m_{\downarrow}^*} - \Delta.$$

Here, spin-up and spin-down states couple to remote odd-parity bands with different energy separations via second-order perturbation, leading to different effective-mass corrections from k·p coupling.

Figure 4 compares the calculated optical absorption spectra (**E**//*ab*, black curves) and excitonic structure (exciton binding energy vs. exciton energy, colored dots) with and without SOC for the $\alpha$, $\beta$, and $\gamma$ monolayers of $WSi_2P_4$. The vertical black dashed line indicates the direct quasiparticle gap $E_g$ at K. The eigenvalues of the excitonic states are shown in the narrow horizontal boxes atop each panel, color-coded by their optical dipole matrix elements (brightness). Note that for 2D materials, the absolute value of the calculated $\varepsilon_2$ depends on the unit-cell volume and thus has no absolute significance. Without SOC, all three phases exhibit a single prominent exciton peak (Peak 1) at 0.51, 0.69, and 0.85 eV for $\alpha$, $\beta$, and $\gamma$, respectively, with comparable brightness. This peak comprises two degenerate bright excitonic states (the first and second excitons), both originating from electron–hole pairs at the K valley. The corresponding exciton binding energies are 0.62, 0.66, and 0.68 eV, respectively. The similar binding energies indicate comparable Coulomb screening in the three phases.

With SOC included, the first and second excitonic states in all three phases become dark, and Peak 1 splits into two: Peak A and Peak B. The lower-energy Peak A appears at 0.31 eV (binding energy 0.54 eV), 0.47 eV (0.63 eV), and 0.64 eV (0.67 eV) for $\alpha$, $\beta$, and $\gamma$, respectively, while Peak B appears at 0.69 eV (0.59 eV), 0.87 eV (0.46 eV), and 0.97 eV (0.66 eV). Peak B comprises multiple excitonic states; we quote the binding energy of the brightest one as representative. The $\alpha$ phase exhibits the strongest Peak A, followed by $\beta$, with $\gamma$ the weakest. Furthermore, in $\alpha$ and $\beta$, Peak A is brighter than Peak B, whereas in γ, Peak B is brighter than Peak A.

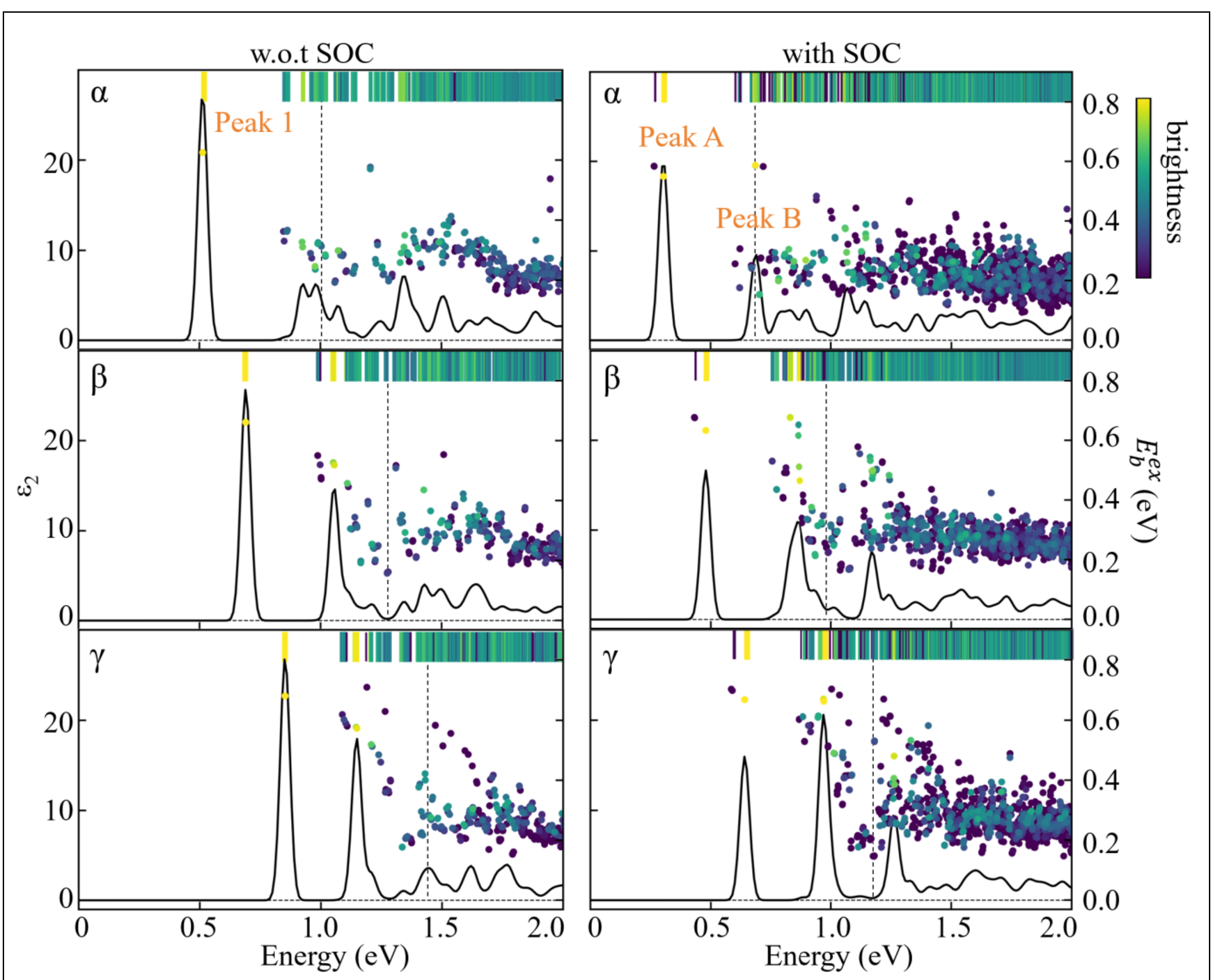

Figure 4. Optical absorption spectra (**E**//*ab*, black curves) and exciton binding energies $E_b^{ex}$ (dots) without (left panels) SOC and with SOC (right panels) of three $WSi_2P_4$ monolayers. The vertical black dashed line is the position of QP band gap ($E_g$) at K. An energy broadening of 0.05 eV is used in the calculation of $\varepsilon_2$. The excitonic state eigenvalues are shown in the horizontal boxes on top of the main plots. The eigenvalues and binding energies are color-coded with their dipole matrix elements (represent the brightness).

To understand these brightness patterns, we correlate them with the conduction-band splitting $\Delta_{CB}$ at K discussed above. In the $\alpha$ phase, $c_1$ is spin-down and $v_1$ is also spin-down near K, so the $v_1 \rightarrow c_1$ transition is allowed, as evidenced by the sharp peak in the transition dipole moment at K (Figure 3b). This spin-allowed radiative channel provides a fingerprint for understanding why Peak A in $\alpha$ is the brightest among the three phases and significantly stronger than Peak B (primarily from $v_2 \rightarrow c_1$). For the $\beta$ phase, $\Delta_{CB}$ = 0.5 meV is very small, placing the band structure near K in a near-degenerate regime. This near-degeneracy makes the energy difference between the $v_1 \rightarrow c_1$ and $v_1 \rightarrow c_2$ transitions very small, rendering Peak A relatively bright as well. For the $\gamma$ phase, $\Delta_{CB}$ = 24.0 meV is significantly larger than in the $\alpha$ and $\beta$ phases, so Peak A (primarily $v_1 \rightarrow c_2$) is less bright than Peak B (primarily $v_2 \rightarrow c_1$), as confirmed in Figure 5.

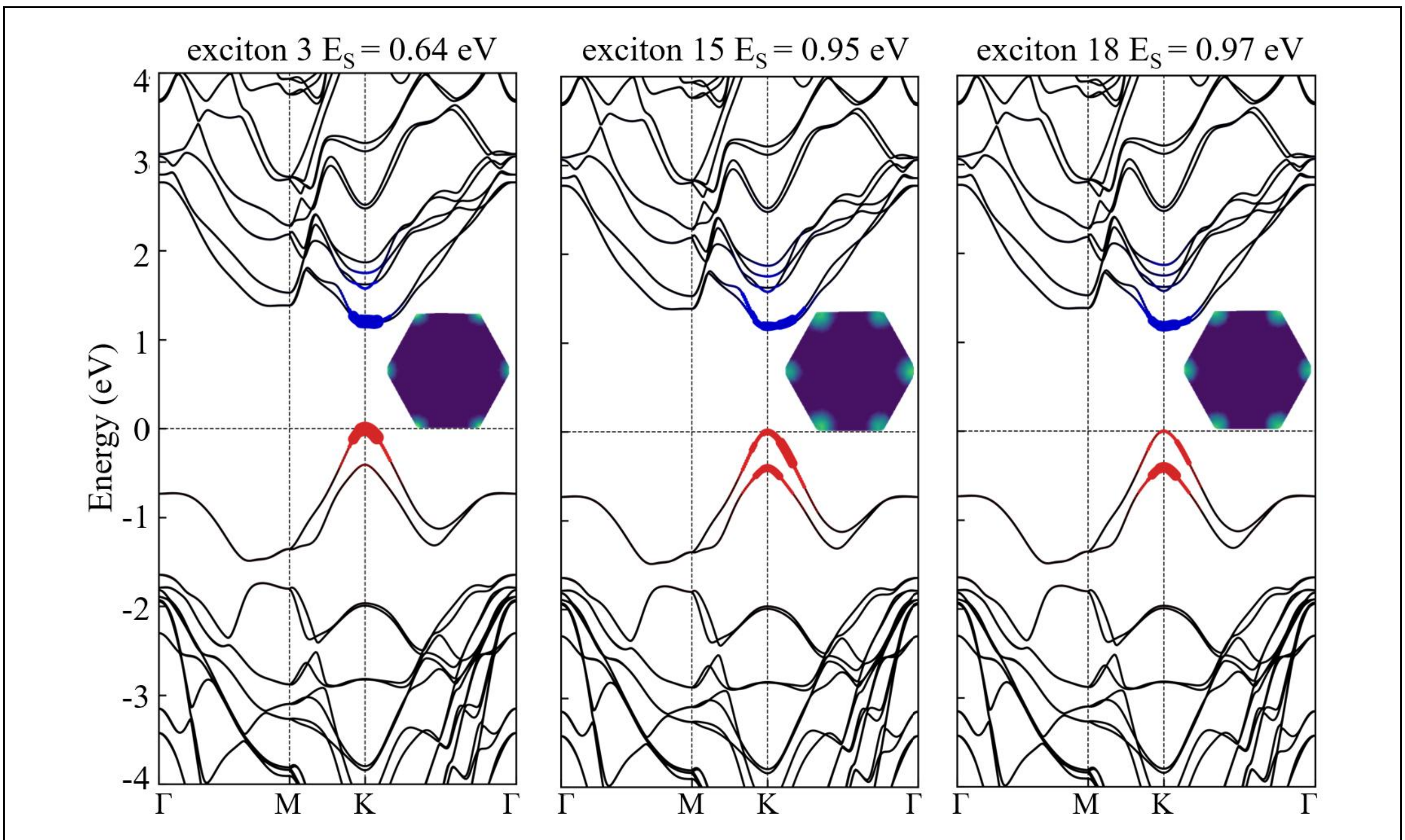

Figure 5. The band- and $k$-resolved electron ($|A^S_{c\boldsymbol{k}}|^2$) and hole ($|A^S_{v\boldsymbol{k}}|^2$) amplitudes of selected excitons with an excitation energy $E^S$ = 0.64 eV (Peak A) and excitation energies with $E^S$ = 0.95 eV and $E^S$ = 0.97 eV (Peak B) of γ-phase $WSi_2P_4$ monolayer. Inset is the $k$-resolved pair ($|A^S_{\boldsymbol{k}}|^2$) amplitude.

To further elucidate the excitonic structure, we analyze the excitonic wavefunctions in detail. Figure 5 shows the band- and $k$-resolved electron $|A^S_{v\boldsymbol{k}}|^2$ and hole $|A^S_{c\boldsymbol{k}}|^2$ amplitudes[37] for three representative excitonic states of $\gamma$-$WSi_2P_4$. Exciton 3 (left panel, corresponding to Peak A) has a binding energy of ~0.67 eV, while excitons 15 and 18 (middle and right panels, corresponding to Peak B) have binding energies of 0.61 and 0.66 eV, respectively. The electron and hole amplitudes of these excitons are highly concentrated in a small region of the Brillouin zone near K. The hole state of Peak A (bright exciton 3) originates from the $v_1$ band in all three phases. However, the electron state differs among the phases. Detailed analysis (Figure S5) reveals that in $\alpha$, the electron state of exciton 3 is a mixture of $c_1$ and $c_2$. Since $c_2$ is spin-matched with $v_1$ (except the tiny band crossing region), it supports bright exciton formation. In contrast, $c_1$ is spin-mismatched with $v_1$ (except the tiny band crossing region), and the resulting triplet excitons are typically dark. It is important to note that exciton brightness is determined by the symmetry of the total exciton wavefunction and electron–hole exchange interactions, not solely by single-particle spins. Exchange interaction strongly mixes singlet and triplet states.[38-39] Although the single-particle bands at K are pure spin states (protected by PST), the exciton wavefunction is a coherent superposition of electron and hole states over a finite momentum range near K. For $\beta$ and $\gamma$, the electron state of exciton 3 originates almost entirely from $c_2$, which is spin-matched with $v_1$. However, because the $c_2$–$c_1$ splitting is significantly larger in γ than in β, the brightness of exciton 3 in $\gamma$ is reduced relative to $\beta$. Finally, Peak B exhibits pronounced mixed character (Figure S6), consistent with its composition from multiple excitonic states. Neglecting nonradiative channels, photoexcited electrons rapidly relax to the CBM. In the *β* and *γ* phases, the K valley is protected by PST. Since $c_1$ and $v_1$ are spin-mismatched, the excited-state lifetimes in these phases are expected to be long.[40] In contrast, the *α* phase exhibits a band crossing in the CBM, creating a tiny region near K where $c_1$ and $v_1$ are spin-matched. This region can open a radiative channel, making *α*-$WSi_2P_4$ a promising candidate for high-quantum-efficiency emitters.

We have performed a comprehensive GW-BSE study of direct-bandgap monolayer $WSi_2P_4$, revealing its polymorph-dependent spin texture, band splitting, and excitonic properties. The three phases—*α*, *β*, and *γ*—exhibit distinct spin textures governed by their point-group symmetries. The $D_{3h}$ *α* and *γ* phases exhibit a PST protected by the $\sigma_h$ mirror plane, weakly modulated near Γ by *k*-mediated interband SOC mixing, while the polar $C_{3v}$ *β* phase exhibits Rashba-type spin splitting. SOC splits the main absorption peak into two peaks whose relative brightness is controlled by the conduction-band splitting $\Delta_{CB}$ at K. The *α* phase, with its tiny band crossing of opposite-spin branches, exhibits the brightest low-energy exciton owing to a spin-allowed radiative channel. These results establish the $WSi_2P_4$ as a platform that provides a foundation for future studies on direct-gap, strong-SOC $MSi_2X_4$ systems for quantum optoelectronic applications.

**Supporting Information**

The Supporting Information is available free of charge at https://...

(I) PBE band structure of $WSi_2P_4$ without SOC. (II) PBAND. (III) Band spin splitting around Γ. (IV) Definition of the *k*-resolved exciton amplitudes.

**Notes**

The authors declare no competing financial interest.

**ACKNOWLEDGMENTS**

F. J. acknowledges the support of Zhejiang Provincial Natural Science Foundation of China (LQN25A040011), National Natural Science Foundation of China (12504264), and the Fundamental Research Funds for the Provincial Universities of Zhejiang (KYS075624288). K. W. and G. X. acknowledges the support of National Natural Science Foundation of China (No. 12504439, 12104118). S. X. thanks the supported by the Opening Project of State Key Laboratory of Functional Crystals and Devices (SKLFCD2026B05). Y. W. acknowledge the support from National Natural Science Foundation of China (12574265) and Youth S&T Talent Support Programme of Guangdong Provincial Association for Science and Technology (GDSTA) (2026QNRC034).

**DATA AVAILABILITY**

The data that support the findings of this article can be acquired by contacting the corresponding author.

**REFERENCE**

1. Kim, K. S.; Kwon, J.; Ryu, H.; Kim, C.; Kim, H.; Lee, E.-K.; Lee, D.; Seo, S.; Han, N. M.; Suh, J. M.; Kim, J.; Song, M.-K.; Lee, S.; Seol, M.; Kim, J., The future of two-dimensional semiconductors beyond

Moore's law. *Nature Nanotechnology* **2024,** *19* (7), 895-906.

2. Turunen, M.; Brotons-Gisbert, M.; Dai, Y.; Wang, Y.; Scerri, E.; Bonato, C.; Jöns, K. D.; Sun, Z.; Gerardot, B. D., Quantum photonics with layered 2D materials. *Nature Reviews Physics* **2022,** *4* (4), 219-236.
3. Hong, Y.-L.; Liu, Z.; Wang, L.; Zhou, T.; Ma, W.; Xu, C.; Feng, S.; Chen, L.; Chen, M.-L.; Sun, D.-M.; Chen, X.-Q.; Cheng, H.-M.; Ren, W., Chemical vapor deposition of layered two-dimensional MoSi2N4 materials. *Science* **2020,** *369* (6504), 670-674.
4. Latychevskaia, T.; Bandurin, D. A.; Novoselov, K. S., A new family of septuple-layer 2D materials of MoSi2N4-like crystals. *Nature Reviews Physics* **2024,** *6* (7), 426-438.
5. Zhou, T.; Xu, C.; Ren, W., The van der Waals MoSi2N4 materials family. *Nature Reviews Materials* **2025,** *10* (12), 907-928.
6. Wu, Y.; Tang, Z.; Xia, W.; Gao, W.; Jia, F.; Zhang, Y.; Zhu, W.; Zhang, W.; Zhang, P., Prediction of protected band edge states and dielectric tunable quasiparticle and excitonic properties of monolayer MoSi2N4. *npj Computational Materials* **2022,** *8* (1), 129.
7. Wang, L.; Shi, Y.; Liu, M.; Zhang, A.; Hong, Y.-L.; Li, R.; Gao, Q.; Chen, M.; Ren, W.; Cheng, H.-M., Intercalated architecture of MA2Z4 family layered van der Waals materials with emerging topological, magnetic and superconducting properties. *Nature Communications* **2021,** *12* (1), 2361.
8. Sun, S.; Xu, C.; Ji, K.; Xia, Y.; Tong, B.; Tong, J.; Chen, C.; Zhao, W.; Zhou, D.; Wang, Q.; Yang, L.; Wang, Q.; Li, M.; Zheng, B.; Ma, L.-P.; Liu, X.; Liu, Z.; Zhu, M.; Liu, K.; Liu, P.; Jiang, K.; Cheng, H.-M.; Ren, W., Wafer-scale growth of highly stable p-type semiconducting monolayer MoSi2N4 single crystals. *Nature Materials* **2026,** *25* (9), 1540-1548.
9. Li, S.; Wu, W.; Feng, X.; Guan, S.; Feng, W.; Yao, Y.; Yang, S. A., Valley-dependent properties of monolayer MoSi2N4, WSi2N4, and MoSi2As4. *Physical Review B* **2020,** *102* (23), 235435.
10. Sun, M.; Re Fiorentin, M.; Schwingenschlögl, U.; Palummo, M., Excitons and light-emission in semiconducting MoSi2X4 two-dimensional materials. *npj 2D Materials and Applications* **2022,** *6* (1), 81.
11. Yang, C.; Song, Z.; Sun, X.; Lu, J., Valley pseudospin in monolayer MoSi2N4 and MoSi2As4. *Physical Review B* **2021,** *103* (3), 035308.
12. Yang, Y.; Wang, X.; Sun, L.; Jia, F.; Ruan, Y.; Feng, T.; Wu, Y., Tailoring the many-body effects and phase configurations in monolayer MSi2X4 (M = Mo, W; X = N, P, As, Sb) for wide-range bandgap engineering. *Physical Chemistry Chemical Physics* **2026,** *28* (14), 8856-8863.
13. Woźniak, T.; Faria Junior, P. E.; Ramzan, M. S.; Kuc, A. B., Electronic and excitonic properties of MSi2Z4 monolayers. *Small* **2023,** *19* (19), 2206444.
14. Wang, Q. H.; Kalantar-Zadeh, K.; Kis, A.; Coleman, J. N.; Strano, M. S., Electronics and optoelectronics of two-dimensional transition metal dichalcogenides. *Nature nanotechnology* **2012,** *7* (11), 699-712.
15. Xu, X.; Yao, W.; Xiao, D.; Heinz, T. F., Spin and pseudospins in layered transition metal dichalcogenides. *Nature Physics* **2014,** *10* (5), 343-350.
16. Sheoran, S.; Monga, S.; Phutela, A.; Bhattacharya, S., Coupled Spin-Valley, Rashba Effect, and Hidden Spin Polarization in WSi2N4 Family. *The Journal of Physical Chemistry Letters* **2023,** *14* (6), 1494-1503.
17. Tao, L.; Tsymbal, E. Y., Persistent spin texture enforced by symmetry. *Nature communications* **2018,** *9* (1), 2763.
18. Lu, X.-Z.; Rondinelli, J. M., Discovery principles and materials for symmetry-protected persistent spin textures with long spin lifetimes. *Matter* **2020,** *3* (4), 1211-1225.
19. Jia, F.; Hu, S.; Xu, S.; Gao, H.; Zhao, G.; Barone, P.; Stroppa, A.; Ren, W., Persistent spin-texture

and ferroelectric polarization in 2D hybrid perovskite benzylammonium lead-halide. *The Journal of Physical Chemistry Letters* **2020,** *11* (13), 5177-5183.

20. Manchon, A.; Koo, H. C.; Nitta, J.; Frolov, S. M.; Duine, R. A., New perspectives for Rashba spin–orbit coupling. *Nature materials* **2015,** *14* (9), 871-882.

21. Zheng, F.; Tan, L. Z.; Liu, S.; Rappe, A. M., Rashba spin–orbit coupling enhanced carrier lifetime in CH3NH3PbI3. *Nano letters* **2015,** *15* (12), 7794-7800.

22. Zhang, X.; Shen, J.-X.; Van de Walle, C. G., Three-dimensional spin texture in hybrid perovskites and its impact on optical transitions. *The journal of physical chemistry letters* **2018,** *9* (11), 2903-2908.

23. Qiu, D. Y.; Da Jornada, F. H.; Louie, S. G., Optical spectrum of MoS2: many-body effects and diversity of exciton states. *Physical Review Letters* **2013,** *111* (21), 216805.

24. Kresse, G.; Furthmüller, J., Efficient iterative schemes for ab initio total-energy calculations using a plane-wave basis set. *Physical review B* **1996,** *54* (16), 11169.

25. Hybertsen, M. S.; Louie, S. G., Electron correlation in semiconductors and insulators: Band gaps and quasiparticle energies. *Phys Rev B* **1986,** *34* (8), 5390-5413.

26. Rohlfing, M.; Louie, S. G., Electron-hole excitations and optical spectra from first principles. *Phys Rev B* **2000,** *62* (8), 4927-4944.

27. Deslippe, J.; Samsonidze, G.; Strubbe, D. A.; Jain, M.; Cohen, M. L.; Louie, S. G., BerkeleyGW: A massively parallel computer package for the calculation of the quasiparticle and optical properties of materials and nanostructures. *Comput Phys Commun* **2012,** *183* (6), 1269-1289.

28. Giannozzi, P.; Baroni, S.; Bonini, N.; Calandra, M.; Car, R.; Cavazzoni, C.; Ceresoli, D.; Chiarotti, G. L.; Cococcioni, M.; Dabo, I., QUANTUM ESPRESSO: a modular and open-source software project for quantum simulations of materials. *Journal of physics: Condensed matter* **2009,** *21* (39), 395502.

29. Hamann, D. R., Optimized norm-conserving Vanderbilt pseudopotentials. *Physical Review B—Condensed Matter and Materials Physics* **2013,** *88* (8), 085117.

30. Van Setten, M. J.; Giantomassi, M.; Bousquet, E.; Verstraete, M. J.; Hamann, D. R.; Gonze, X.; Rignanese, G.-M., The PseudoDojo: Training and grading a 85 element optimized norm-conserving pseudopotential table. *Computer Physics Communications* **2018,** *226*, 39-54.

31. Hu, T.; Jia, F.; Zhao, G.; Wu, J.; Stroppa, A.; Ren, W., Intrinsic and anisotropic Rashba spin splitting in Janus transition-metal dichalcogenide monolayers. *Physical Review B* **2018,** *97* (23), 235404.

32. Xiao, D.; Liu, G.-B.; Feng, W.; Xu, X.; Yao, W., Coupled spin and valley physics in monolayers of MoS2 and other group-VI dichalcogenides. *Physical Review Letters* **2012,** *108* (19), 196802.

33. Bentmann, H.; Abdelouahed, S.; Mulazzi, M.; Henk, J.; Reinert, F., Direct Observation of Interband Spin-Orbit Coupling in a Two-Dimensional Electron System. *Physical Review Letters* **2012,** *108* (19), 196801.

34. Schliemann, J., Colloquium: Persistent spin textures in semiconductor nanostructures. *Reviews of Modern Physics* **2017,** *89* (1), 011001.

35. Rashba, E., Properties of semiconductors with an extremum loop. I. Cyclotron and combinational resonance in a magnetic field perpendicular to the plane of the loop. *Sov. Phys.-Solid State* **1960,** *2*, 1109.

36. Zhang, X.; Liu, Q.; Luo, J.-W.; Freeman, A. J.; Zunger, A., Hidden spin polarization in inversion-symmetric bulk crystals. *Nature Physics* **2014,** *10* (5), 387-393.

37. Tang, Z.; Cruz, G. J.; Jia, F.; Wu, Y.; Xia, W.; Zhang, P., Stacking up Electron-Rich and Electron-Deficient Monolayers to Achieve Extraordinary Mid-to Far-Infrared Excitonic Absorption: Interlayer Excitons in the C3B/C3N Bilayer. *Physical Review Applied* **2023,** *19* (4), 044085.

38. Becker, M. A.; Vaxenburg, R.; Nedelcu, G.; Sercel, P. C.; Shabaev, A.; Mehl, M. J.; Michopoulos, J. G.; Lambrakos, S. G.; Bernstein, N.; Lyons, J. L., Bright triplet excitons in caesium lead halide perovskites.

*Nature* **2018,** *553* (7687), 189-193.
39. Bir, G. L.; Pikus, G. E., Symmetry and strain-induced effects in semiconductors. *Wiley* **1974**.
40. Yuan, J.; Wei, Q.; Sun, M.; Yan, X.; Cai, Y.; Shen, L.; Schwingenschlögl, U., Protected valley states and generation of valley-and spin-polarized current in monolayer MA 2 Z 4. *Physical Review B* **2022,** *105* (19), 195151.